\documentclass[english]{article}
\usepackage[T1]{fontenc}
\usepackage[latin9]{inputenc}
\usepackage[a4paper]{geometry}
\usepackage{url}
\usepackage{amsmath}
\usepackage{amssymb}
\usepackage{graphicx}
\usepackage{setspace}
\usepackage{cite}
\usepackage[unicode=true]{hyperref}



\begin{document}
\title{Scaling of variability measures in hierarchical demographic data}
\author{Aleksejus Kononovicius${}^{1,}$\thanks{email: \protect\href{mailto:aleksejus.kononovicius@tfai.vu.lt}{aleksejus.kononovicius@tfai.vu.lt};
website: \protect\url{https://kononovicius.lt}}, Justas Kvedaravicius${}^2$}
\date{\small${}^1$Institute of Theoretical Physics and Astronomy, Vilnius
University\\${}^2$Faculty of Physics, Vilnius University}
\maketitle

\begin{abstract}
Demographic heterogeneity is often studied through the geographical
lens. Therefore it is considered at a predetermined spatial resolution,
which is a suitable choice to understand scalefull phenomena. Spatial
autocorrelation indices are well established for this purpose. Yet
complex systems are often scale-free, and thus studying the scaling
behavior of demographic heterogeneity may provide valuable insights.
Furthermore, migration processes are not necessarily influenced by
the physical landscape, which is accounted for by the spatial autocorrelation
indices. The migration process may be more influenced by the socio-economic
landscape, which is better reflected by the hierarchical demographic
data. Here we explore the scaling behavior of variability measures
in the United Kingdom 2011 census data set. As expected, all of the
considered variability measures decrease as the hierarchical scale
becomes coarser. Though the non--monotonicity is observed, it can
be explained by accounting for the imperfect hierarchical relationships.
We show that the scaling behavior of variability measures can be qualitatively
understood in terms of Schelling's segregation model and Kawasaki--Ising
model.
\end{abstract}

\section{Introduction}

While the focus of sociophysics is the opinion dynamics \cite{Castellano2009RevModPhys},
the lack of theoretical models comparable to the empirical data drives
the need to understand the spatial heterogeneity of the people and
their opinions \cite{FernandezGracia2014PRL,Vazquez2022Entropy}.
Most established models in sociophysics describe the temporal evolution
of opinions and hand--wave the spatial dynamics in favor of running
simulations on social networks \cite{Castellano2009RevModPhys,Vazquez2022Entropy}.
The main two ways one could introduce spatial dynamics is by considering
opinion formation on a highly sophisticated social network with embedded
spatial information \cite{FernandezGracia2014PRL}, or one could use
the ergodic property of many opinion dynamics models to assume that
different spatial units are simply independent realizations of the
same opinion formation process \cite{Sano2016,Kononovicius2017Complexity,Braha2017PlosOne}.
The first approach is highly sophisticated and requires huge amounts
of data for proper model calibration, the second approach is too simplistic
in its assumption of the absence of spatial interdependencies. Recently
\cite{Kononovicius2019CompJStat} proposed a novel approach that could
allow a reasonably simple and general approach to the model of spatial
demographic and electoral heterogeneity. Understanding how the demographic
variability changes across the different scales of hierarchical demographic
data is one of the steps towards this goal.

\section{Scaling behavior in the UK 2011 census data}

Here, we explore the United Kingdom 2011 census data set\footnote{Publicly available from \url{https://www.nomisweb.co.uk} website.}.
Our goal is to understand the randomness in the observed demographic
heterogeneity qualitatively. Spatial autocorrelation indices (such
as Moran's~I or Geary's~C) are typically used for such purpose \cite{Cliff1981Pion},
but these indices are applicable when spatial data is available. Here,
we consider census data across various urban hierarchical scales and
ignore any explicitly spatial features.

The first step in our analysis is to convert the raw numbers $X_{i}^{\left(k\right)}$
(number of people belonging to the $k$-th demographic category living
in $i$-th spatial unit) into fractions:
\begin{equation}
x_{i}^{\left(k\right)}=\frac{X_{i}^{\left(k\right)}}{\sum_{m}X_{i}^{\left(m\right)}}.\label{eq:define-frac}
\end{equation}
The sum in the denominator represents the total number of people living
in the $i$-th unit (it goes over all complementary demographic categories).
Next, we estimate the demographic heterogeneity by measuring the variability
of $x_{i}^{\left(k\right)}$. The heterogeneity is measured in this
way for every hierarchical scale considered. Finally, the measured
variability values are normalized so that all measures would be equal
to unity at the finest scale.

\begin{figure}[htp]
\begin{centering}
\includegraphics[width=0.8\textwidth]{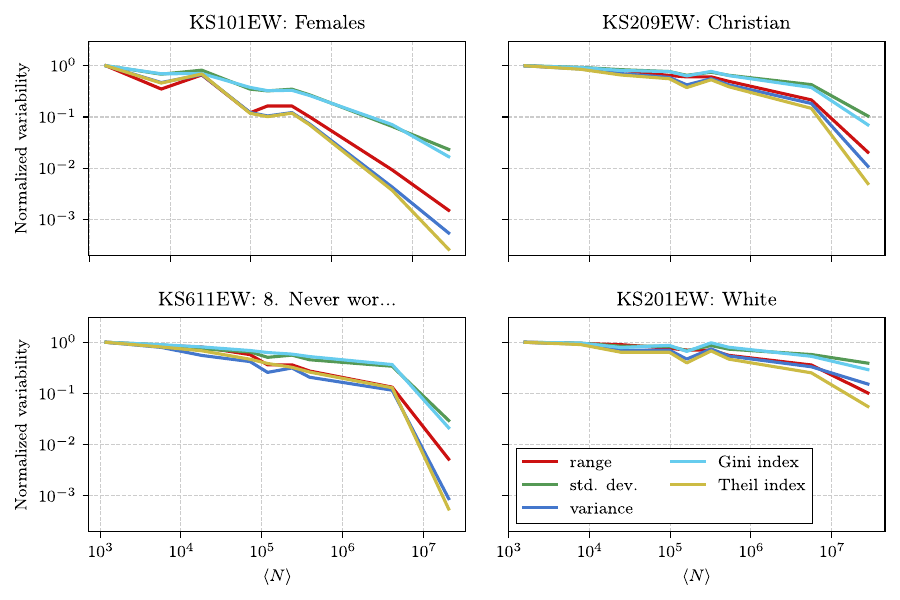}
\par\end{centering}
\caption{Normalized variability (colored curves) of four different demographic
categorizations across different hierarchical scales.\label{fig:normalized-variability}}
\end{figure}

As can be seen in Fig.~\ref{fig:normalized-variability}, all of
the considered variability measures exhibit the same qualitative behavior:
they decay as the scales become coarser. Some of the measures decay
faster than others, but in general, the measures either scale together
with the standard deviation or with the variance, which indicates
that different scaling behavior in measures is just a matter of units.
Therefore, further in this paper we will consider only standard deviation
as the measure of variability (heterogeneity).

Comparison across the different demographic categorizations yields
desired qualitative insights: in some cases normalized variability
decays faster than in others. The faster decay indicates that the
population becomes more uniform faster, which is expected if the population
is not segregated with respect to that categorization. There is little
evidence for segregation based on sex, but on the other hand, religious,
ethnic, and economic segregation is a well--documented phenomenon
\cite{Oka2019EEP}.

\section{Baseline null model}

To provide a baseline for the understanding of the randomness in the
observed demographic heterogeneity let us introduce a null model.
The core idea of the null model would be to remove any spatial or
hierarchical associations present in the empirical data.

To this end, the input of the null model is the data from the finest
scale and the desired number of units at a coarser scale. The finest
scale units are randomly assigned to groups, the number of which is
equal to the desired number of units at the coarser scale. Random
assignment happens with one important restriction: the groups need
to have as similar number of units assigned to them as possible. Units
within the groups are then combined to produce the randomized coarser
scale units. To avoid Simpson's paradox, we add the raw numbers $X_{i}^{\left(k\right)}$
instead of simply averaging the fractions $x_{i}^{\left(k\right)}$.
Repeating these steps for all coarser scales allows us to see how
fast the normalized variability decays if the process is purely random.

\begin{figure}[htp]
\begin{centering}
\includegraphics[width=0.7\textwidth]{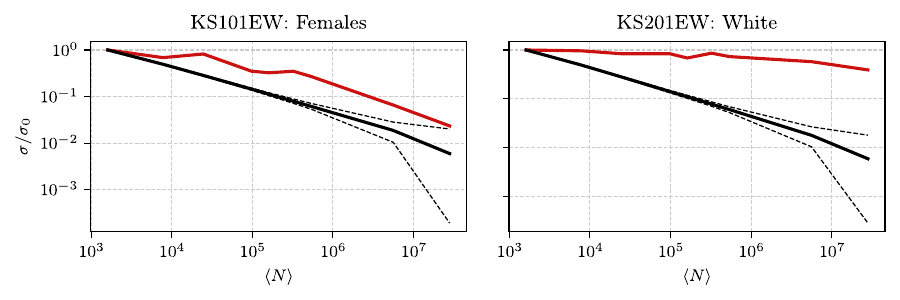}
\par\end{centering}
\caption{Normalized standard deviation of the two different demographic categorizations
across the different scales: empirical data (red curve) compared against
the null model (median is shown as solid black curve, while $95\%$
confidence interval is limited by the thin dashed black curves).\label{fig:null-model}}
\end{figure}

As can be seen in Fig.~\ref{fig:null-model}, both sex and race curves
show higher variability than would be expected if the migration and
demographic change were a random processes (the empirical curves are
above the null model curve). Notably, the null model in both cases
produces power--law scaling behavior
\begin{equation}
\frac{\sigma}{\sigma_{0}}\sim\frac{1}{\sqrt{\left\langle N\right\rangle }}.\label{eq:scaling-rel}
\end{equation}

Such scaling relationship is expected, as when increasing the scale
we are effectively averaging values on the lowest scale. Then the
variance of the mean of $N$ random values $\bar{x}=\frac{1}{N}\sum_{i=1}^{N}x_{i}$
would be given by:
\begin{align}
\mathrm{Var}\left[\bar{x}\right] & =\mathbb{E}\left[\left(\bar{x}-\mathbb{E}\left[\bar{x}\right]\right)^{2}\right]
 =\mathbb{E}\left[\left(\frac{1}{N}\sum_{i=1}^{N}x_{i}-\mathbb{E}\left[\frac{1}{N}\sum_{i=1}^{N}x_{i}\right]\right)^{2}\right] 
 =\frac{1}{N^{2}}\mathbb{E}\left[\left(\sum_{i=1}^{N}x_{i}-\mathbb{E}\left[\sum_{i=1}^{N}x_{i}\right]\right)^{2}\right]=\nonumber \\
 & =\frac{1}{N^{2}}\mathrm{Var}\left[\sum_{i=1}^{N}x_{i}\right]
 =\frac{1}{N^{2}}\left\{ \sum_{i=1}^{N}\mathrm{Var}\left[x_{i}\right]+\sum_{i=1}^{N}\sum_{j=1}^{N}\mathrm{Cov}\left(x_{i},x_{j}\right)\right\} .
\end{align}
For the independent random values $x_{i}$, with $\mathrm{Cov}\left(x_{i},x_{j}\right)=0$,
we have:
\begin{equation}
\sigma=\sqrt{\mathrm{Var}\left[\bar{x}\right]}=\sqrt{\frac{1}{N}\mathrm{Var}\left[x_{i}\right]}=\frac{\sigma_{0}}{\sqrt{N}}.
\end{equation}
This relationship will hold for any underlying distribution of the
random values on the lowest scale. Though, in the empirical data we
would expect that these values would follow Beta distribution or a
mixture of Beta distributions \cite{Kononovicius2019CompJStat}. In
Figs.~\ref{fig:normalized-variability}~and~\ref{fig:null-model}
we have $\left\langle N\right\rangle $ as independent variable, because
in the empirical data set higher hierarchical scales are produced
by combining not necessarily the same number of smaller units.

For the ethnic data, finding the slower decay of variability (in comparison
to the null model) was expected. But for the sex data, the obtained
slower decay is unexpected, though the difference from the null model
is relatively small. The deviations in the latter case could likely
be explained by the data anonymization procedures and the presence
of institutional installations favoring one of the sexes (e.g., military
bases, convents).

\section{Non--monotonicity of variability}

Note that the empirical curves in Figs.~\ref{fig:normalized-variability}
and \ref{fig:null-model} are non--monotonic, while the intuition
from the baseline null model would suggest that the variability measures
should monotonically decrease with coarser scales. The non--monotonicity
cannot be explained by the objective socio--demographic mechanisms
such as segregation, but instead can be replicated by assuming that
it is an observational effect. Namely, non--monotonicity can be replicated
if we allow for the hierarchical associations to be imperfect: meaning
that the coarser scale units are not necessarily produced by combining
the finer scale units. For example, some finer scale units might be
split and then be incorporated into different coarser scale units.
If this is allowed then the variability will decay non--monotonically.

\begin{figure}
\begin{centering}
\includegraphics[width=0.7\textwidth]{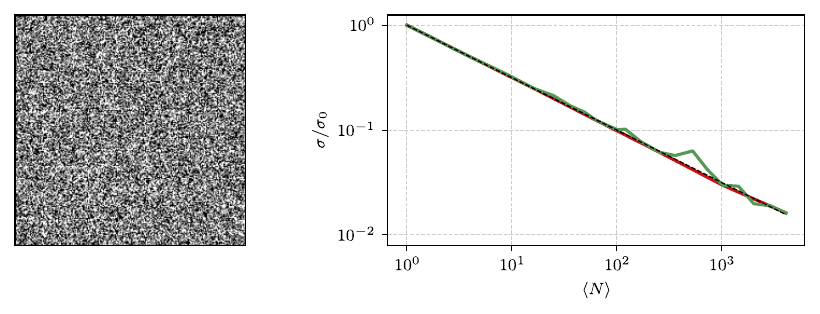}
\par\end{centering}
\caption{Random grid (left) and normalized standard deviation curve (right)
measured across the consistent (red curve) and inconsistent hierarchical
scales (green curve). Thin black dashed curve shows Eq.~\eqref{eq:scaling-rel}.\label{fig:splits}}
\end{figure}

To verify this intuition by numerical simulation, let us generate
a random grid with $256\times256$ cells. Let each cell, with equal
probability, contain either a value of $0$ or a value of $1$. Then,
let us measure the variability of such grid at different scales (up
to $4\times4$ grid, meaning that the cells at the coarsest scale
are $64\times64$ in size). If the cells are combined consistently
(without splitting finer scale units), then the obtained curve is
monotonic (the red curve in Fig.~\ref{fig:splits}). On the other
hand, if the cells are combined inconsistently (i.e., allowing the
splits), then the non--monotonicity can be observed (the green curve
in Fig.~\ref{fig:splits}).

\section{Modeling (anti--)segregation}

\begin{figure}
\begin{centering}
\includegraphics[width=0.7\textwidth]{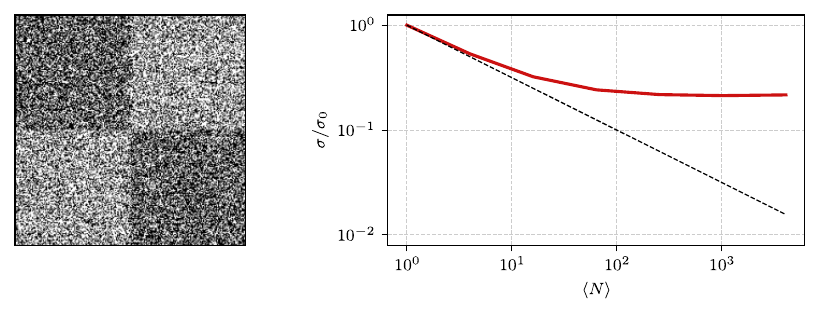}
\par\end{centering}
\caption{Segregated random grid (left) and normalized standard deviation curve
(right) measured across the consistent scales (red curve). Thin black
dashed curve shows Eq.~\eqref{eq:scaling-rel}.\label{fig:segregation}}
\end{figure}

One of the simplest models for segregation would be to split the grid
into regions with different probabilities for a cell on a grid to
contain $0$ or $1$, thus artificially creating spatial divisions.
One such grid of size $256\times256$ is shown in Fig.~\ref{fig:segregation}.
It was obtained with $p_{1}=0.6$ in two $128x128$ regions (these
regions appear darker in the figure), and symmetrically $p_{1}=0.4$
in the other two $128\times128$ regions (these appear lighter). The
corresponding normalized standard deviation curve (also shown in Fig.~\ref{fig:segregation})
indicates that at the lowest scales the heterogeneity (as measured
by standard deviation) scales consistently with the random model.
Saturation at the higher scales is induced by the way we have introduced
segregation.

\begin{figure}[htp]
\begin{centering}
\includegraphics[width=0.4\textwidth]{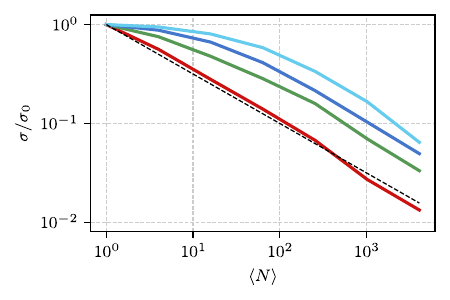}
\par\end{centering}
\caption{Normalized standard deviation curves obtained for the Schelling's
model with different happiness thresholds: $0$ (red curve), $25\%$
(green), $37.5\%$ (blue), and $50\%$ (cyan). All of the obtained
curves are above or overlap the thin black dashed curve, which corresponds
to the random grid model. \label{fig:segregation-schelling}}
\end{figure}

\begin{figure}[htp]
\begin{centering}
\includegraphics[width=0.4\textwidth]{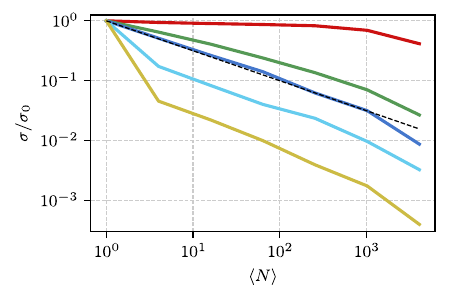}
\par\end{centering}
\caption{Normalized standard deviation curves obtained for the Kawasaki--Ising
model. The curves above or overlapping the thin black dashed curve
(corresponding to the random grid model) were obtained in ferromagnetic
regime for different inverse temperatures: $\beta=2$ (red curve),
$1$ (green), $0.1$ (blue). The curves below were obtained in anti--ferromagnetic
regime for $\beta=2$ (cyan), $4$ (yellow).\label{fig:segregation-ising}}
\end{figure}

Notably, the empirical normalized standard deviation curves have slightly
different shape than predicted by the segregated random grid model.
Namely, the empirical curves exhibit slower decay of heterogeneity
at the lowest scales, with the decay of the curves becoming faster
as the scale increases. This shape can be qualitatively replicated
by Schelling's segregation model \cite{Hatna2012JASSS} (see Fig.~\ref{fig:segregation-schelling}),
or the Kawasaki--Ising model \cite{Kawasaki1966PR,Kawasaki1966PR2}
(see Fig.~\ref{fig:segregation-ising}). While the classic formulation
of the Schelling's segregation model doesn't exhibit anti--segregation
regime (though it can be easily modified to do so), but anti--segregation
regime can be found in the Kawasaki--Ising model with anti--ferromagnetic
interactions. Both models were simulated on $128\times128$ grid,
being initialized with a random configuration with a constraint that
there should be the same number of cells containing $0$ and $1$
($16\%$ of cells were empty in simulations with the Schelling's model).
In both cases the state was observed after $10^{7}$ Monte Carlo steps.

\section{Conclusions}

Here we have explored scaling of variability measures in hierarchical
demographic data. From the empirical point-of-view we have examined
United Kingdom 2011 data set. We have considered five different variability
measures, and found that they mostly exhibit the same qualitative
behavior. Therefore to carry out the analysis choosing either one
of them would be sufficient. While the range is the easiest variability
measure to calculate, it is highly sensitive to outliers, thus it
is less suitable than the variance or the standard deviation. On the
other hand Gini and Theil indices require more computational effort
to calculate than the variance or the standard deviation. As the standard
deviation has the same units as the quantity it is calculated from,
we prefer it over the variance.

In the empirical analysis we have observed non--monotonicity of the
normalized variability curves, but as all the measures produce qualitatively
the same non--monotonic behavior, we are prompted to look at other
observational imperfections. We were able to reproduce the non--monotonicity
by assuming that hierarchical relationships are imperfect: that some
of the smallest scale units are getting split when producing higher
scale units.

Finally, we have shown that the obtained empirical curves cannot be
well explained by artificially introducing segregation into the random
grid model. Instead a bit more complicated model, such as the Schelling's
segregation model or the Kawasaki--Ising model are necessary to reproduce
qualitative behavior of the empirical curves.

In the foreseeable future we expect to use this preliminary analysis
as a basis of producing hierarchical correlation index similar in
purpose to the spatial correlation indices. And to further enhance
spatial models of opinion dynamics (such as \cite{Kononovicius2019CompJStat}).

\begin{singlespace}
\section*{Author contributions}

{\bfseries AK}: Conceptualization, Methodology, Software, Writing,
Visualization. {\bfseries JK}: Conceptualization, Methodology.


\end{singlespace}

\end{document}